\documentclass[aps,pra,amsmath,amsfonts,mathtools,amssymb,twocolumn,groupedaddress]{revtex4-1}
  
\usepackage[]{graphicx} 
\usepackage[]{amsfonts}
\usepackage{ulem}
\usepackage[usenames,dvipsnames]{color}

\begin{document}


\title{Master equation for multilevel interference in a superradiant medium}


\author{Aleksei Konovalov}
\author{Giovanna Morigi}
\affiliation{Theoretische Physik, Saarland University, 66123 Saarbr\"ucken, Germany}



\date{\today}

\begin{abstract}
We derive a master equation for a superradiant medium which includes multilevel interference betwen the individual scatterers. The derivation relies on the Born-Markov approximation and implements the coarse graining formalism. The master equation fulfils the Lindblad form and contains terms describing multilevel interference {between parallel transitions of a single atom}, multi-atom interference between identical transitions, and multi-atom interference between different electronic transitions with parallel dipoles.  This formalism is then applied to determine the excitation spectrum of two emitters using the parameters of the Hydrogen transitions 2S$_{1/2}\,\to$4P$_{1/2}$ and 2S$_{1/2}\,\to$4P$_{3/2}$, where the gap between the parallel dipoles is of the order of GHz. The distortion of the signal due to the interplay of multilevel and multi-emitter interference is analysed as a function of their distance. These results suggest that interference between parallel dipolar transition can significantly affect the spectroscopic properties of optically dense media.  
\end{abstract}


\maketitle

\section{Introduction}

Superradiance generally denotes a phenomenon which enhances radiation. In quantum optics, it originates from quantum interference in the light emission by an ensemble of atoms, molecules, or other types of resonant emitters which form an optically-dense medium \cite{Dicke:1954,GrossHaroche:1982,Friedberg:1972}. In free space this requires that the average interparticle distance is smaller than the wavelength of the scattered radiation. Then, the coupling of the individual atomic transitions with the modes of the electromagnetic field can be effectively described in terms of collective dipoles and the radiative properties depend on the collective spin quantum numbers \cite{Dicke:1954}. Superradiant (and subradiant) scattering plays a relevant role in the spectroscopy of dense atomic gases \cite{Grimes:2016,Pellegrino:2014,Araujo:2016,Bromley:2016,Corman:2017,Peyrot:2018,Jennewein:2018}, it could enhance transport of light in organic semiconductors \cite{Orgiu:2015}, and it is the key mechanism of recent realizations of ultranarrow lasers \cite{Meiser:2009,Norcia:2016}.

 \begin{figure}
 	\includegraphics[width=1\linewidth]{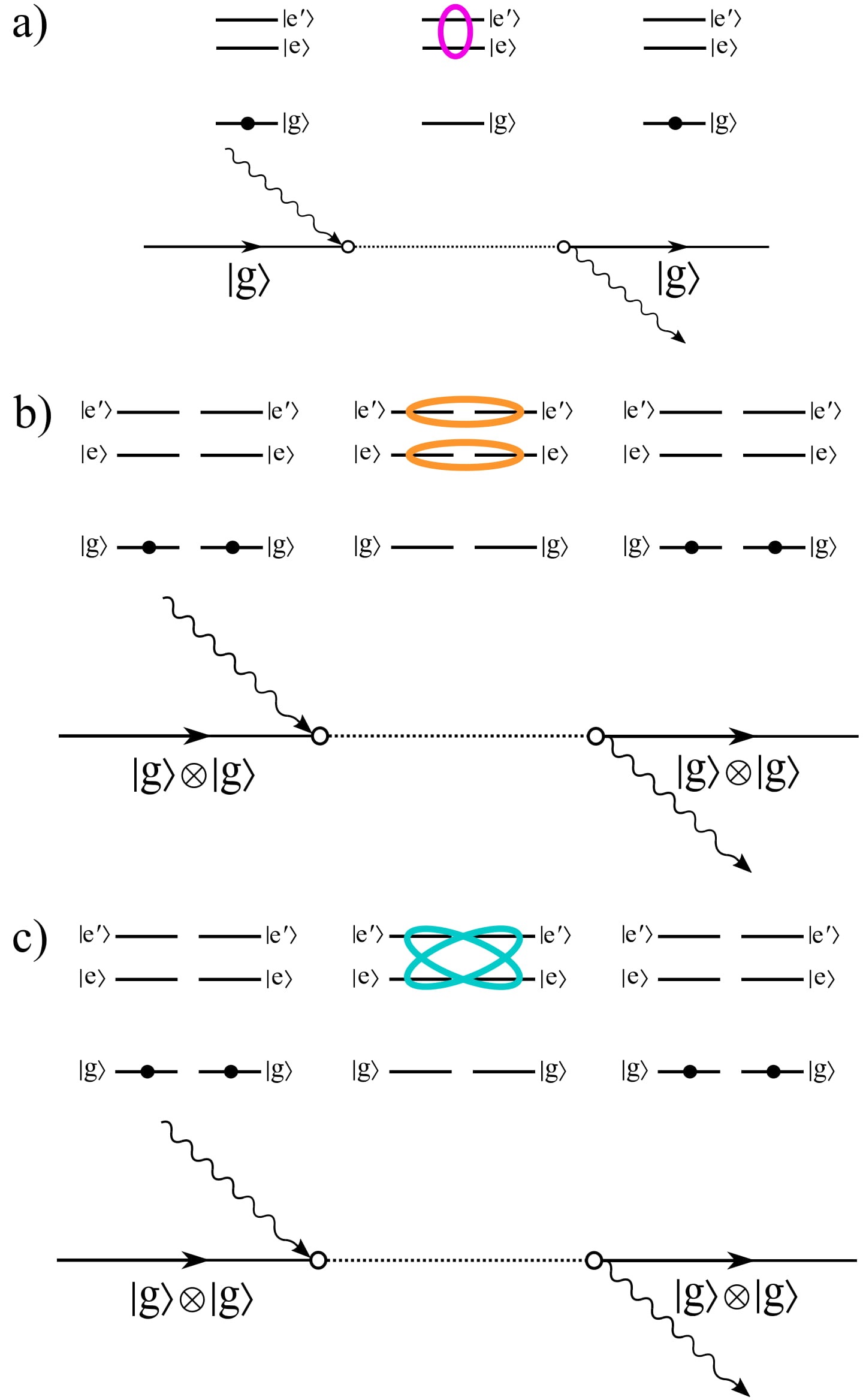}
 	\caption{\label{Fig:1} Interfering processes leading to photon scattering by resonant emitters. The emitter's relevant states are the ground state  $|g\rangle$ and the excited states $|e\rangle$,  $|e'\rangle$, the transitions $|g\rangle\to |e\rangle$ and $|g\rangle\to |e'\rangle$ have parallel dipole moments. The horizontal lines sketch the scattering processes, the wavy lines the photon, the level schemes give the corresponding occupation of the emitters' internal levels (dot and circles). The emitters are initially in the ground state (solid line). Photon absorption (first wiggle line) can excite a coherent superposition of (a) the excited states of a single emitter, (b) the resonant states of the two emitters, and (c) different excited states of the two emitters but with parallel dipoles. Photon emission (second wiggle line) projects the emitters in the same final state. In this work we analyse the spectroscopic features due to the interference of these three processes. }  		
 \end{figure}   

Superradiant light scattering is often described by means of a perturbative expansion in the atom-photon interactions and using the Born-Markov approximation \cite{Agarwal,GrossHaroche:1982,Friedberg:1972,Lehmberg:1970,Milonni:1974,James:1993,Carmichael,Fleischhauer:1999,Yelin:Review,Breuer}. Most theoretical treatments focus on two-level dipolar transitions \cite{Agarwal,GrossHaroche:1982,Friedberg:1972,Lehmberg:1970,Milonni:1974,James:1993,Carmichael,Fleischhauer:1999,Lin:2012,Zhu:2016}, some also including the possible degeneracy of the ground or excited state of the transition \cite{James:1993,Lin:2012,Zhu:2016}. These treatments successfully predict experimental measurements at sufficiently low optical densities. Qualitative discrepancies have been found when comparing the predictions of these models with recent experiments with dense atomic media \cite{Pellegrino:2014,Bromley:2016,Jennewein:2018}. This requires one to assess the effects of terms which are typically discarded or only partially considered.  

{In this work we derive a master equation for an optically dense medium and set our focus on vacuum-induced interference \cite{Milonni:1976,Cardimona:1983,Ficek,Kiffner:2010}. Vacuum-induced interference refers to interference phenomena between electronic transitions coupled to common modes of the electromagnetic field. If the transitions are dipolar, they are denoted by parallel dipoles. Interference occurs also when the electromagnetic field modes are in the vacuum, and is qualitatively different from laser-induced interference \cite{Berman:1998}. In closed level structures these effects can be tested by means of quantum beat spectroscopy and are expected to give rise to "steady-state quantum beats" \cite{Cardimona:1983,Ficek}. They are also expected to play an important role in high-precision spectroscopy \cite{Horbatsch,Yost:2014,Buchheit:2016,Udem:2019}. In this work we determine the Born-Markov master equation of multilevel scatterers in an optically dense medium and which includes interference terms between parallel dipoles }. For this purpose we derive the master equation by applying the coarse-graining formalism of Ref. \cite{Lidar:2001,Majenz:2013}. The master equation we obtain preserves the Lindblad form and in the limit of one single emitter it reduces to the coarse-grained master equation of Ref. \cite{Buchheit:2016}. We then apply it to determine the excitation spectrum and the light shift of two identical emitters, each composed by two parallel dipoles sharing the same ground state. 
In this simplified model we show that collective scattering results from the coherent sum of three processes, which we illustrate in Fig. \ref{Fig:1}: (a) the interference between parallel dipoles of the individual atoms, (b) the interference between resonant transitions of different atoms, and (c) the interference between parallel dipoles of different atoms. Here, we argue that in an optically dense medium they can give rise to measurable shifts of the spectroscopic lines.  

This work is organized as follows. In Sec. \ref{chapter:derivation} we present the derivation of the master equation by eliminating the degrees of freedom of the electromagnetic field within the Born-Markov approximation and by implementing the coarse-graining method developed in Ref. \cite{Lidar:2001}. By these means we obtain a superoperator that fulfils the Lindblad form. This superoperator consistently describes interference processes between parallel dipoles of the individual atoms and interference processes of different atoms. In Sec. \ref{chapter:results} we then consider the specific example of two emitters, composed by two parallel dipoles sharing the same ground state, and determine their excitation spectrum using the parameters of the transitions 2S$_{1/2}\to$4P$_{1/2}$ and 2S$_{1/2}\to$4P$_{3/2}$ of Hydrogen atom. By means of a simple fitting function we argue that the interference effects give rise to measurable shifts of the resonance lines. Finally, in Sec. \ref{chapter:conclusions} we draw the conclusions and discuss outlooks of this work. The appendices contain details of the calculations in Sec. \ref{chapter:derivation} and Sec. \ref{chapter:results}.
   
\section{Derivation of The Superradiant Master equation}
\label{chapter:derivation}

In this Section we report the derivation of the Born-Markov master equation for an optically dense atomic or molecular medium. Our derivation follows the lines of textbook derivations \cite{Milonni:1976,GrossHaroche:1982, Carmichael,Breuer}, and extend it by implementing the coarse-grained method developed in Ref. \cite{Lidar:2001}. This allows us to take systematically into account the interference of parallel dipoles and at the same time to preserve the Lindblad form of the master equation. In the single-atom limit our master equation reproduces the one derived in Ref. \cite{Buchheit:2016}, which includes the interference processes between parallel dipoles in a single atom.

For convenience, in the following we assume an ensemble of emitters with identical electronic transitions. This formalism, nevertheless, can be straightforwardly extended  to ensembles of different particles (which could also be a mixture of atoms and molecules) with quasi-resonant transitions. The relevant assumption is that the emitters are pinned at given positions and are distinguishable particles. Our starting point is the von-Neumann equation governing the coherent dynamics. Below we provide the salient steps leading to the corresponding coarse-grained master equation for the emitters' internal degrees of freedom. 

\subsection{Multilevel emitters interacting with the quantum electromagnetic field}

We consider $N$ emitters interacting with the modes of the electromagnetic field (EMF) in the volume $V$. We assume that the particles are pinned at the positions $\vec R_\alpha$, with $\alpha=1,\ldots,N$. We denote by $\mathcal H$ the Hilbert space of the emitters' internal degrees of freedom and of the EMF's degrees of freedom,  $\mathcal H= \mathcal H_A\otimes  \mathcal H_R$. The time evolution of the density matrix $\hat{\chi}(t)$, describing the state of photons and emitters, is governed by the von-Neumann equation
\begin{equation}   \label{eq:von_neumann}
\partial_t \hat{\chi} = [\hat H,\hat{\chi}]/i\hbar\,,
\end{equation}   
where $\hat H$ is the Hamiltonian determining the dynamics, which we decompose into the sum of the Hamiltonian $\hat{H}_{\text{A}}$ for the emitters' (internal) degrees of freedom, the Hamiltonian $\hat{H}_{\text{R}}$ for the free EMF, and the emitter-photon interactions $\hat{V}$:
\begin{equation} \label{eq:gen_ham}
\hat{H}=\hat{H}_{\text{A}}+\hat{H}_{\text{R}}+\hat{V}\,.
\end{equation}  
We remark here that $\hat{H}_{\text{R}}\equiv \hat 1_A\otimes \hat{H}_{\text{R}}$ and  $\hat{H}_{\text{A}}\equiv\hat{H}_{\text{A}}\otimes \hat 1_R$, 
where $\hat 1_R$ and by $\hat 1_A$ the identity operators in the Hilbert spaces $\mathcal{H}_{\text{R}}$ and $ \mathcal{H}_{\text{A}}$, respectively. Thus, we use the same notation for the operator $\hat H_{j=A,R}$ defined in the extended Hilbert space  $\mathcal H$ and in the reduced Hilbert space $\mathcal H_j$

{\it The emitters' Hamiltonian.} The emitters Hamiltonian describes the dynamics of the internal degrees of freedom of $N$ emitters: 
 $$\hat{H}_{\text{A}} =\sum_{\alpha=1}^N \hat{H}_{{\text{A}}_{\alpha}}\,, $$
 where $\hat{H}_{{\text{A}}_{\alpha}}$ is the Hamiltonian of emitter $\alpha=1,\ldots,N$ at position $\vec R_\alpha$ and we assume that the size of the center-of-mass wavepacket is much smaller than the interparticle distance (in Eq. \eqref{H:alpha} we omit to explicitly write that $\hat{H}_{{\text{A}}_{\alpha}}$ is the identity operator in the Hilbert space of the emitters with $\beta\neq \alpha$).  
 We consider here only the lowest electronic bound states assuming that the system is at room temperature. The spectrum of each emitter is discrete and the Hamiltonian in diagonal form reads
\begin{equation}
\label{H:alpha}
\hat{H}_{{\text{A}}_{\alpha}} =\sum_n E_{n} |n \rangle_{\alpha} \langle n |\,,
\end{equation}
with $E_n$ the eigenvalue and $|n \rangle_\alpha$ the corresponding eigenvector for the emitter at the position $\vec R_\alpha$. In a more general treatment, where the emitters might not be identical and/or in the presence of spatial inhomogeneity, then the energy also depends on the label $\alpha$. 

{\it The quantum electromagnetic field.} We treat the EMF in second quantization and choose the Coulomb gauge. We denote the quantization volume by $V$ and assume periodic boundary conditions. The energy of the field relative to the vacuum energy reads:
\begin{equation}
\label{H:EMF}
\hat{H}_{\text{R}}= \sum_{\lambda} \hbar\omega_{\lambda} \hat{a}_{\lambda}^{\dagger}\hat{a}_{\lambda}\,,
\end{equation}
where $\lambda$ denotes the sum over the EMF modes  and the sum has an upper cutoff given by the energy $\hbar\omega_{\text{cutoff}} \sim mc^2$, with $m$ the electron mass.  The modes are here traveling waves and are fully characterized by the wave vector $\vec{k}_{\lambda}$ and by the transverse polarization $\vec{e}_{\lambda}$, with the frequency $\omega_\lambda=c|\vec{k}_\lambda|$ and $c$ the speed of light in vacuum. Operators $\hat{a}_{\lambda}$ and $\hat{a}_{\lambda}^{\dagger}$ annihilate and create, respectively, a photon of mode $\lambda$, and fulfil the bosonic commutation relations $[\hat{a}_{\lambda},\hat{a}_{\lambda'}^{\dagger}]=\delta_{\lambda,\lambda'}$ and  $[\hat{a}_{\lambda},\hat{a}_{\lambda'}]=0$.

The initial state of the EMF field is assumed to be given by the thermal distribution 
\begin{equation}
\hat{R}=\exp{(-\hat{H}_{\text{R}}/k_BT)}/Z\,, 
\label{eq:R}
\end{equation}
where $k_B$ is Boltzmann's constant, $T$ is the temperature, and $Z=\text{Tr}\{\exp{(-\hat{H}_{\text{R}}/k_BT)}\}$ is the partition function. Within the validity of the Born approximation, $\hat{R}$ gives the state of the EMF at all times. Here we assume room temperatures, $T \sim 300 \, \text{K}$.

{\it Emitter-photon interactions.} Emitter-photon interactions are here treated in the electric-dipole approximation. Operator $\hat{V}$ is the sum of the interactions of the fields with each emitter, $\hat{V}=\sum_{\alpha=1}^N \hat{V}_{\alpha} $, with
\begin{equation}    
\label{eq:interaction_op}
\hat{V}_{\alpha}=\hbar\sum_{n}\hat{\Gamma}_{n}^{\alpha}\hat{\sigma}_{n}^{\alpha}\,,  
\end{equation}
where the sum is over all pairs of electronic levels $n=(n_1,n_2)$ coupled by an electric dipole transition. 
Here, operator $\hat{\sigma}_{n}^{\alpha}$ describes the transition between $|n_1\rangle_\alpha$ and $|n_2\rangle_\alpha$:
$$\hat{\sigma}_{n}^{\alpha}\equiv |n_1\rangle_\alpha\langle n_2|\,.$$
The corresponding coupling strength is determined by the coupling operator $\hat{\Gamma}_{n}^{\alpha}$, which acts over the degrees of freedom of the electromagnetic field and reads:
\begin{equation}     \label{eq:coupling_op} 
\hat{\Gamma}_{n}^{\alpha}= \sum_{\lambda} \left(g_{n}^{\alpha\lambda}\hat{a}_{\lambda}e^{i\vec{k}_{\lambda}\vec{R}_{\alpha}} + 
\bar{g}_{n}^{\alpha\lambda}\hat{a}^{\dagger}_{\lambda}e^{-i\vec{k}_{\lambda}\vec{R}_{\alpha}}\right)\,.
\end{equation}
The coupling strengths $g_{n}^{\alpha\lambda}$ have the dimensions of a frequency and are below given in Gauss units and in the length gauge:
\begin{eqnarray}
&& g_{n}^{\alpha\lambda}=-i\sqrt{\frac{2\pi \omega_{\lambda}}{\hbar V}}\,\vec{d}_{n}^{\alpha}\cdot\vec{e}_{\lambda}\,,\\
&&\bar{g}_{n}^{\alpha\lambda}=i\sqrt{\frac{2\pi \omega_{\lambda}}{\hbar V}}\,\vec{d}_{n}^{\alpha}\cdot(\vec{e}_{\lambda})^*\,,
\end{eqnarray}
with $\vec{d}_{n}^{\alpha}$ the dipole moment of the transition, which is the matrix element of the dipole operator $\hat{\vec{d}}_\alpha$ and reads $\vec{d}_{n}^{\alpha}={}_\alpha\langle n_1|\hat{\vec{d}}_\alpha|n_2\rangle_\alpha$. We remark that this description applies the long-wave approximation, and thus it is valid when the size of the electronic wave packet is smaller than the optical wavelength. {Moreover, in our model we did not include the self-energy which appears in the length gauge (see Refs. \cite{Cohen-Tannoudij,Rubio:2018} for an insightful discussion). }

For later convenience we introduce the frequency $\omega_{n}$:
\begin{equation}
\omega_n=(E_{n_1}-E_{n_2})/\hbar\,.
\end{equation}
By definition it can also take negative values.

\subsection{Master equation for an ensemble of multilevel emitters}

We now proceed in deriving the Born-Markov master equation using the coarse-grained formalism. The procedure repeats in the essential steps the one of Ref.\
\cite{Buchheit:2016}, with some notable differences due to the many-body nature of the problem.

We first introduce the density matrix $\hat{\rho}(t)$ describing the state of the emitters at time $t$. Operator $\hat{\rho}(t)$ is defined in the Hilbert space $\mathcal H_{\rm A}$ and is related to the density matrix $\hat\chi(t)$ by the equation: $\hat{\rho}(t)={\rm Tr}_R\{\hat\chi(t)\}$, where ${\rm Tr}_R$ denotes the partial trace over the degrees of freedom of the EMF. 

We now consider the von-Neumann equation, Eq. \eqref{eq:von_neumann}, and move to the interaction picture with respect to Hamiltonian $\hat H_0=\hat H_{\rm A}+\hat H_{\rm R}$. 
We denote the system's density matrix in interaction picture by
\begin{equation}
\tilde\chi(t)=\hat U_0(t)^\dagger \hat{\chi}(t)\hat U_0(t)\,,
\end{equation}
where we have introduced the unitary operator $\hat U_0(t)=\exp(\hat H_0t/({\rm i}\hbar))$. In this representation the reduced density matrix of the system is related to the reduced density matrix in Schr\"odinger picture by the relation:
$$\tilde{\rho}(t)={\rm Tr}_R\{\tilde\chi(t)\}={\rm e}^{-\hat H_{\rm A}t/({\rm i}\hbar)}\hat{\rho}(t){\rm e}^{\hat H_{\rm A}t/({\rm i}\hbar)}\,.$$

In interaction picture the unitary operator determining the  time evolution reads:
\begin{equation} \label{eq:time_evol}
{\tilde{U}}(t,t') = \mathcal{T} \exp \left(-\frac{\rm i}{\hbar} \int_{t}^{t'} dt_1 {\tilde{V}}(t_1)  \right)\,,
\end{equation} 
where ${\tilde{V}}(t)=\hat U_0(t)^\dagger \hat{V}\hat U_0(t)$ and $\mathcal T$ denotes the time ordering, such that 
$$\mathcal{T} {\tilde{V}}(t_1) {\tilde{V}}(t_2) = {\tilde{V}}(t_1) {\tilde{V}}(t_2) \theta (t_1-t_2) + {\tilde{V}}(t_2) {\tilde{V}}(t_1) \theta (t_2-t_1)\,,
$$ with $\theta (t)$ the Heaviside function. Using this formalism, at $t'>t$ the state $\hat{ \tilde{\chi}}(t)$ evolves into state 
\begin{equation}
\label{eq:dm} 
\tilde\chi(t')= {\tilde{U}}(t,t') \tilde\chi(t) {\tilde{U}}(t,t')^{\dagger} \,.
\end{equation}

\subsubsection{Dyson equation and Born-Markov approximation} 

Let now $\Delta t=t'-t>0$ denote a finite and sufficiently small time step, which we quantify later. We write the Dyson series of the right-hand side of Eq. \eqref{eq:dm} till second order in the interaction, but keep the exact form. After tracing out the EMF degrees of freedom we obtain the expression
\begin{eqnarray} \label{eq:decomposition}
&&\tilde{\rho}(t+\Delta t) =
  \tilde{\rho}(t) +\Delta t \sum_{\alpha}\Lambda_{1}^\alpha\tilde{\rho}(t)\\
&&+\Delta t \sum_{\alpha,\beta}\frac{1}{\Delta t}\int_{t-\Delta t}^{t+\Delta t}{\rm d}T\int_{-\Delta t}^{\Delta t}{\rm d}\tau \theta(\tau)\Lambda_2^{\alpha,\beta}(T,\tau)\tilde{\rho}(T-\tau)\,.\nonumber
\end{eqnarray}
The terms $\Lambda_1^\alpha$, $\Lambda_2^\alpha$ on the RHS are linear maps, the subscript indicate the order in the Dyson expansion. In deriving Eq. \eqref{eq:decomposition} we have made the Born approximation at the initial time $t$, namely, we have assumed that there are no quantum correlations at time $t$ between EMF and emitter. This corresponds to writing $\tilde{\chi}(t)=\hat R\otimes  \tilde{\rho}(t)$ where here $\hat R$ is the thermal state of the EMF, Eq. \eqref{eq:R}. 

 The map $\Lambda_1^\alpha$ acts over the Hilbert space of the emitter $\alpha$ and is given by:
\begin{eqnarray}
 \Lambda_1^\alpha\tilde{\rho}(t)&=&\frac{1}{{\rm i}\hbar\Delta t} \int_{t}^{t+\Delta t} dt_1 \text{Tr}_{\text{R}}\left\{\left[{\tilde{V}}_{\alpha}(t_1) , \tilde\chi(t)\right] \right\} \,.\\
                                                   &=&\frac{1}{{\rm i}\hbar}\left[ \langle{\tilde{V}}_{\alpha}(t)\rangle_\text{R},\tilde{\rho}(t)\right]\,,\nonumber 
\end{eqnarray}
where between the first and the second line we have applied the Born approximation and introduced the time-averaged operator (here in interaction picture):
\begin{equation}
\label{V:alpha}
\langle{\tilde{V}}_{\alpha}(t)\rangle_\text{R}=\frac{1}{\Delta t} \int_{t}^{t+\Delta t} \text{Tr}_{\text{R}}\left\{{\tilde{V}}_{\alpha}(t_1) \hat R \right\}\,.
\end{equation}
Note that operator $\hat V_\alpha$, Eq. \eqref{eq:interaction_op}, vanishes over the thermal state of the EMF, Eq. \eqref{eq:R}. 
The second integrand of Eq. \eqref{eq:decomposition} contains the Heaviside function $\theta(\tau)$ and includes also the coupling between different emitters. Its detailed form is reported in Appendix \ref{App:MEq}.

Equation \eqref{eq:decomposition} is generally valid for sufficiently short time intervals $\Delta t$, over which one can assume that the Born approximation holds. After some time, in fact, the interactions establish quantum correlations between system and reservoir. These correlations can be neglected when the interactions can be treated perturbatively. 

The master equation becomes local in time when the Markov approximation holds. The Markov approximation consists in approximating $\tilde{\rho}(T-\tau)\approx \tilde{\rho}(t)$ in Eq. \eqref{eq:map:0}. {It is equivalent to the Wigner-Weisskopf approximation for the propagator \cite{Cohen-Tannoudij} } and is justified when the characteristic time scale $\tau_R$ of the correlation function $C_{\alpha\beta}(\tau)$, Eq. \eqref{eq:correlation_function}, is orders of magnitude smaller than the system's relaxation time. In a thermal bath the correlation function is composed by a term which decays exponentially with the correlation time $\tau_R=\hbar/k_BT $ and by power-law tails that can be discarded for typical evolution times \cite{Cohen-Tannoudij,Ingold}. At room temperatures, $T \sim 300$ K, this time is of the order of $\tau_R\sim 10^{-13}$ sec. This time shall be compared with the relaxation time of the system. For optical transitions the natural linewidth of a single atom, $\gamma\sim 2\pi\times 10^6-10^8$ Hz,  fulfils $\gamma\tau_R\ll 1$. In this limit we can choose the time scale $\Delta t$ such that $\tau_R\ll \Delta t\ll 1/\gamma$ and ignore memory effects in the integral. 

In the presence of dipole-dipole interactions there are some issues to be considered: in first place, superradiance gives rise to an $N$-fold enhancement of the single atom decay rate, thus when $N\gamma$ becomes comparable with $1/\tau_R$ the approximation becomes invalid. This is the regime where one can observe the Dicke phase transition in an ensemble of two-level systems \cite{Hepp}, and where the assumptions at the basis of this treatment break down. At the same time, subradiant states can be characterized by extremely small linewidths. Observing their decay requires one to analyse the system's dynamics over long time scales, over which the power-law tails of the correlation function can become important. These considerations suggest that the formalism shall be revisited for media with very high optical dense media. 

\subsubsection{Coarse-grained master equation}

In what follows we assume an optically dense medium for which the Born-Markov approximation is valid. Then, from Eq. \eqref{eq:decomposition} we derive the Born-Markov master equation (now back in 
Schr\"odinger picture):
\begin{equation}
\label{eq:ME}
\partial_t\hat \rho=\frac{1}{{\rm i}\hbar}[\hat H_A+\hat H_S,\hat \rho(t)]+\mathcal{L}_{D}\hat \rho(t)\,,
\end{equation}
where Hamiltonian $\hat H_S$ and superoperator (dissipator) $\mathcal L_D$ contain both the single-atom as well as the interatomic interference terms between parallel dipoles.  The details of the derivation are  standard and are reported in Appendix \ref{App:MEq}. The master equation is valid for any time $t>0$ within a grid whose resolution is determined by the coarse-grained time-scale $\Delta t$. {As a consequence, the coefficients multiplying the terms of the operator $\hat H_S$  and the superoperator $\mathcal L_D$ are scaled by the function}
\begin{equation}
\label{Eq:Theta:ij}
\Theta_{ij}^{(\Delta t)}=\frac{\sin((\omega_i+\omega_j)\Delta t/2)}{(\omega_i+\omega_j)\Delta t/2}\,.
\end{equation}
This term selects transitions which are resonant within the resolution set by the coarse-graining time $\Delta t$. For optical transitions, this factor selects a pair of frequencies $\omega_i$ and $\omega_j$ with opposite signs. Correspondingly, it selects terms in Hamiltonian and dissipator where the pairs of operators $\hat\sigma_i^\alpha\hat\sigma_j^\beta$ describe an excitation and a de-excitation along two (quasi-)resonant transitions. For convenience, we introduce the operator $\hat\zeta_i^{\alpha\dagger}\equiv\hat\sigma_i^{\alpha}$, which describes a transition $i_2\to i_1$ with $\bar\omega_i=\omega_i>0$ and dipole moment $\vec D_i^{\alpha*}=\vec d_i^\alpha$. Then, the operators appearing in the master equation are of the form $\hat\zeta_i^{\alpha\dagger}\hat\zeta_j^{\beta}$ or $\hat\zeta_i^{\alpha}\hat\zeta_j^{\beta\dagger}$ and the factor \eqref{Eq:Theta:ij} now reads
\begin{equation}
\label{Eq:Theta:ij}
\Theta_{ij}^{(\Delta t)}=\frac{\sin((\bar\omega_i\pm \bar\omega_j)\Delta t/2)}{(\bar\omega_i\pm \bar\omega_j)\Delta t/2}\,.
\end{equation} 
 In what follows we discard the processes where two transitions are simultaneously excited or de-excited, corresponding to the $+$ sign in the argument of Eq. \eqref{Eq:Theta:ij}.

{\it Hamilton operator.} The Hamiltonian term due to the interaction with the EMF is given by the expression
$$\hat H_S=\sum_\alpha\langle \hat V_\alpha\rangle_R+\frac{1}{2}\sum_{\alpha,\beta}\hat H_{\alpha\beta}^S\,,$$
where $\langle \hat V_\alpha\rangle_R$ is given in Eq. \eqref{V:alpha} and is now reported in Schr\"odinger picture. This latter term vanishes, since we assume that the EMF is in the thermal state. The Hamilton operator $\hat H_{\alpha\beta}^S$ contains the frequency shifts and couplings due to the multilevel interference, and is derived from the expression (here given in interaction picture):
\begin{eqnarray}
\tilde H_{\alpha\beta}^S&=&-\frac{\rm i}{2\hbar\Delta t}\int_t^{t+\Delta t}{\rm d}t_1\int_t^{t+\Delta t}{\rm d}t_2\\
& &\times\theta(t_1-t_2){\rm Tr}_R\left\{[\tilde V_\alpha(t_1),\tilde V_\beta(t_2)]R(t)\right\}+{\rm H.c.}\,.\nonumber
\label{H:Lamb}
\end{eqnarray}
For $\alpha=\beta$ it is the Hamilton operator for a single atom and it coincides with the operator derived in Ref.
\cite{Buchheit:2016}. For  $\alpha \neq \beta$ 
it describes the Hamiltonian terms due to the dipole-dipole interactions, including the interference between all parallel transitions of different atoms. We report it in the form which includes both cases:
\begin{eqnarray} 
 \label{eq:Lamb+-}
\hat{H}_{\alpha\beta}^S = -\hbar  \sum_{i,j} &&\left[\left(\Delta_{ij}^{\alpha\beta-}+\Delta_{ij}^{\alpha\beta(T)}\right) \hat{\zeta}_{i}^{\alpha\dagger}\hat{\zeta}_{j}^{\beta}   \right.\\
&&+\left.\left(\Delta_{ij}^{\alpha\beta+}-\Delta_{ij}^{\alpha\beta(T)}\right)^*   \hat{\zeta}_{i}^{\alpha}\hat{\zeta}_{j}^{\beta\dagger}\right]+{\rm H.c.} \nonumber\,,
\end{eqnarray}
where $\Delta_{ij}^{\alpha\beta(T)}=\Delta_{ij}^{\alpha\beta-}(T)-\Delta_{ij}^{\alpha\beta+}(T)$ and the individual coefficients read (below in Gauss units):
\begin{eqnarray} 
&&\Delta_{ij}^{\alpha\beta\pm}=\Theta_{ij}^{(\Delta t)} \frac{\vec{D}_{i}^{\alpha*}\cdot\vec{D}_{j}^{\,\beta}}{(2\pi)^2 \hbar c^3} \mathcal{P} \int_{0}^{\omega_{\rm cut}} \frac{d\omega \,\omega^3}{\omega \pm \omega_{ij}} F_{\alpha\beta}^{ij}(\vec{R}_{\alpha\beta})\nonumber   \\
\label{eq:coef:1}\\
&&\Delta_{ij}^{\alpha\beta\pm}(T)=\Theta_{ij}^{(\Delta t)} \frac{\vec{D}_{i}^{\alpha*}\cdot\vec{D}_{j}^{\,\beta}}{(2\pi)^2 \hbar c^3}  \mathcal{P} \int_{0}^{\omega_{\rm cut}} \frac{d\omega \,\omega^3n(\omega,T)}{\omega \pm \omega_{ij}} F_{\alpha\beta}^{ij}(\vec{R}_{\alpha\beta})\,.\nonumber   \\    
\label{eq:coef:2} \end{eqnarray}
Here, $\mathcal{P}$ denotes the Cauchy principal value and $\omega_{\rm cut}$ is the cutoff frequency. The frequency 
$$\omega_{ij}=\frac{\bar\omega_i+\bar\omega_j}{2}$$
is the average between the two transition frequencies, and the coefficient $F_{\alpha\beta}^{ij}(\vec{R}_{\alpha\beta})$ depends also on the distance $\vec{R}_{\alpha\beta}=\vec{R}_{\alpha}-\vec{R}_{\beta}$ between the atoms and on the wave number $k=\omega/c$. It takes the form
\begin{eqnarray} \label{eq:F_diffraction_function}
&&F_{\alpha\beta}^{ij}(\vec{R}_{\alpha\beta}) = 
4\pi \left(  j_0(kR_{\alpha\beta})  \left[1- \frac{(\vec{D}_i^{\alpha} \cdot\vec{R}_{\alpha\beta})^*(\vec{D}_j^{\,\beta} \cdot\vec{R}_{\alpha\beta})}{D_i^{\alpha}D_j^{\beta}R_{\alpha\beta}^2}  \right] \right.
\nonumber\\
&&- \left. \frac{j_1(kR_{\alpha\beta})}{kR_{\alpha\beta}} 
\left[1- \frac{3(\vec{D}_i^{\alpha} \cdot\vec{R}_{\alpha\beta})^*(\vec{D}_j^{\,\beta} \cdot\vec{R}_{\alpha\beta})}{D_i^{\alpha}D_j^{\beta}R_{\alpha\beta}^2}  \right] \right)\,,
\end{eqnarray} 
where we used the notation $D_i^\alpha=|\vec{D}_i^{\alpha}|$ and $R_{\alpha\beta}=|R_{\alpha\beta}|$. Here, $j_0(x)$ and $j_1(x)$ are spherical Bessel functions of the first type \cite{Abramowitz-Stegun}. The dependence on the vector joining the two atoms breaks the spherical symmetry and is at the origin of the anisotropic light emission of superradiance \cite{GrossHaroche:1982}. For the case of one atom, $N=1$, one has $F_{\alpha \alpha}^{ij}(0)= 8\pi/3$ \cite{Carmichael}, and Hamiltonian of Eq. \eqref{eq:Lamb+-}
takes the form of the single-atom Hamiltonian of Ref. \cite{Buchheit:2016}.

{\it Dissipator.} The Lindblad term $\mathcal{L}_{D}$ describes the incoherent processes. It can be decomposed into the sum
\begin{equation}
\label{eq:LD}
\mathcal{L}^{D}\hat \rho(t)=\sum_{\alpha,\beta}\mathcal{L}_D^{\alpha\beta}\hat \rho(t)\,,
\end{equation}
where the terms with $\alpha=\beta$ describe the dissipation of $N$ non-interacting atoms, while the terms with $\alpha\neq\beta$ originate from multiple scattering of resonant photons and vanish when the distance between the atoms exceeds several wavelengths. The individual terms are obtained from the expression in interaction picture
\begin{eqnarray*} 
\tilde{\mathcal{L}}_D^{\alpha\beta}\tilde \rho(t)=\frac{1}{2\hbar^2\Delta t}\int_t^{t+\Delta t}{\rm d}t_1\int_t^{t+\Delta t}{\rm d}t_2{\rm Tr}_R\left\{\mathcal A(t_1,t_2)\right\}\,,
\end{eqnarray*}
where
\begin{eqnarray*} 
\mathcal A(t_1,t_2)&=&2\tilde V_\beta(t_1)[\tilde\rho(t)\otimes \tilde R(t)]\tilde V_\alpha(t_2)\nonumber\\
 & &-[\tilde V_\alpha(t_1)\tilde V_{\beta}(t_2),\tilde\rho(t)\otimes \tilde R(t)]_+\,
\end{eqnarray*}
and $[,]_+$ denotes the anticommutator. After performing the integration and going back to Schr\"odinger picture, the individual terms take the form
\begin{eqnarray}  \label{eq:dissipator}
\mathcal{L}_D^{\alpha\beta}\hat \rho(t)&=&  
 \sum_{i,j} (1+n(\omega_{ij},T))\\
& &\times\left(\frac{{\Gamma_{\alpha\beta}^{ij}}}{2}\left[\hat{{\zeta}}_{j}^{\beta} \hat{{\rho}}(t), \hat{{\zeta}}_{i}^{\alpha\dagger}\right] + \frac{{\Gamma_{\alpha\beta}^{ij}}}{2}\left[\hat{{\zeta}}_{j}^{\beta}, \hat{{\rho}}(t) \hat{{\zeta}}_{i}^{\alpha\dagger}\right] \right) \nonumber\\
&+& \sum_{i,j} n(\omega_{ij},T)\nonumber\\
& &\times\left(\frac{{\Gamma_{\alpha\beta}^{ij*}}}{2}\left[\hat{{\zeta}}_{j}^{\beta\dagger} \hat{{\rho}}(t), \hat{{\zeta}}_{i}^{\alpha}\right] + \frac{{\Gamma_{\alpha\beta}^{ij*}}}{2}\left[\hat{{\zeta}}_{j}^{\beta\dagger}, \hat{{\rho}}(t) \hat{{\zeta}}_{i}^{\alpha}\right] \right)\,, \nonumber
\end{eqnarray}
with the damping coefficients
\begin{equation} \label{eq:Gamma_CG}
	\Gamma_{\alpha\beta}^{ij}=\Theta_{ij}^{(\Delta t)}\frac{\vec{D}_{i}^{\alpha*}\cdot\vec{D}_{j}^{\,\beta} }{{2\pi}\hbar c^{3}} \omega_{ij}^{3} F_{\alpha\beta}^{ij}(k_{ij})  
\end{equation}
and $ k_{ij}=\frac{\omega_{ij}}{c}$. We note that for $i\neq j$ the damping coefficients are different from zero if the scalar product $\vec{D}_{i}^{\alpha*}\cdot\vec{D}_{j}^{\,\beta} \neq 0$. 
Master equation \eqref{eq:ME} fulfils the Lindblad form and take into account the multilevel structure of the quantum emitters. 

\subsubsection{Discussion}

We first review the dynamics that the master equation \eqref{eq:ME} predicts for an very dilute ensemble of emitters ($R_{\alpha\beta}\to\infty$), when it is well approximated by $N$ independent experiments with a single atom. In this case the damping coefficients $\Gamma_{\alpha\alpha}^{ii}$ are the Einstein coefficients of spontaneous emission. For $i\neq j$, instead, the coefficients $\Gamma_{\alpha\alpha}^{ij}$ describe processes where two different transitions with parallel dipoles are simultaneously de-excited. These transitions shall be resonant within the frequency resolution of the coarse graining $1/\Delta t$.  This process, even though incoherent, is a quantum interference between spectral lines \cite{Kiffner:2010,Ficek,Horbatsch,Buchheit:2016}. The corresponding terms have been denoted by cross-damping terms in the literature \cite{Yost:2014,Udem:2019}. These dynamics have a corresponding hermitian component in the Hamiltonian term $\hat H_S^{\alpha\alpha}$. The coefficients include an energy shift of the electronic states due to the vacuum fluctuations, which for the ground state is the non-relativistic Lamb shift, as well as a shift due to thermal fluctuations of the EMF. Vacuum and thermal fluctuations give also rise to an effective coupling between electronic levels with parallel dipoles and quasi-resonant frequencies, the coupling coefficients are given by Eqs. \eqref{eq:coef:1}-\eqref{eq:coef:2} after setting $\alpha=\beta$.  They can be estimated by using the approximate relation \cite{Buchheit:2016}
\begin{equation} \label{eq:CST}
\Delta_{ij}^{\alpha\alpha\pm} \approx \frac{1}{2} (\vec{D}_{i}^{\alpha*}\cdot\vec{D}_{j}^{\,\alpha}) \Theta_{ij}^{(\Delta t)} \left( \frac{1}{|\vec{D}_{i}^{\alpha}|^2}\Delta_{ii}^{\alpha\alpha\pm} + \frac{1}{|\vec{D}_{j}^{\alpha}|^2}\Delta_{jj}^{\alpha\alpha\pm} \right)\,.
\end{equation}   

When the interparticle distances are comparable with the wavelength, namely for $\alpha\neq \beta$, Eq. \eqref{eq:ME} is the master equation for optically dense media which now includes quantum interference between transitions with parallel dipoles. Keeping only the terms with $i=j$ one obtains the master equation discussed in the literature \cite{Friedberg:1972,GrossHaroche:1982,James:1993,Zhu:2016}, where the dissipator gives rise to phenomena such as superradiance and subradiance, while the coherent part describes coherent dipole-dipole interaction, including frequency shifts such as the so-called collective Lamb shift \cite{Friedberg:1972,Scully:2009,Roehlsberger:2011,Peyrot:2018}. Our derivation highlights, in addition, the existence of interference terms between quasi-resonant transitions of different atoms with parallel dipoles both in the incoherent as well as in the coherent part of the master equation. 

{We finally remark that, by taking the limit $\Delta t\to 0$, thus for an infinitesimally small coarse-grained time-scale, the function \eqref{Eq:Theta:ij} becomes a Dirac-delta function. Then, the coarse-grained master equation reduces to the Born-Markov master equation discussed for instance in Refs. \cite{Carmichael,Cohen-Tannoudij,Scully:book}. In this limit, however, one discards effects due to the finite time-scale of the reservoir dynamics, thus interference phenomena between parallel transitions which are close in frequency but not exactly resonant. The coarse-graining master equation allows one to include these dynamics in a systematic way. We refer the interested reader to Refs. \cite{Breuer,Lidar:2001,Buchheit:2016,Farina:2019} for discussions on the coarse-grained master equation and to the next section for a discussion about the choice of $\Delta t$.}
 
 \section{Excitation spectrum of two emitters} \label{chapter:results}
 
 \begin{figure}
  	\includegraphics[width=1\linewidth]{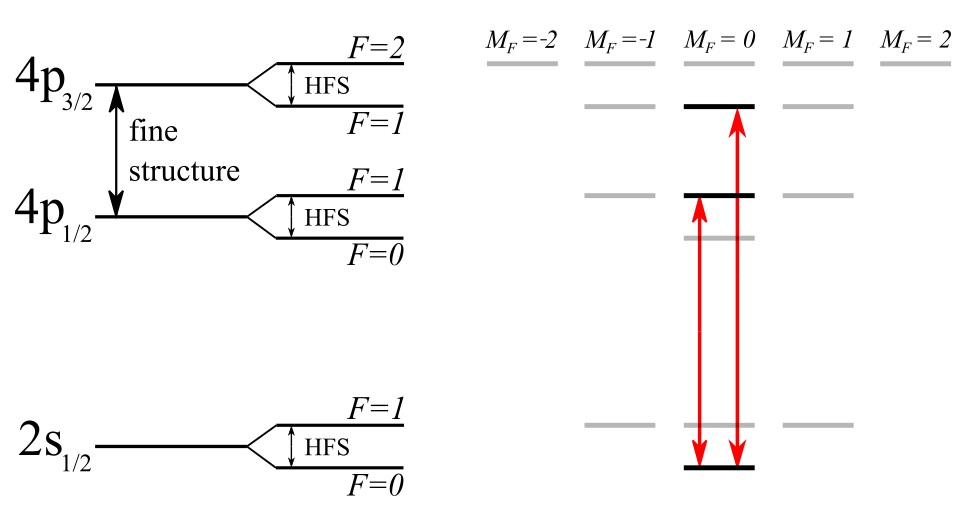}
  	\caption{\label{fig:Level_scheme}
  The structure of electronic states 2S and 4P of the hydrogen atom (left panel) and the transitions we consider in the numerical simulations of this work (right panel): The three states which we consider in this work are marked with the black colour, the scattering transitions are indicated by the red (grey) arrows.}
  \end{figure} 	

We now determine the excitation spectrum of two emitters, which are pinned at the positions $\vec R_1=0$ and $\vec R_2=\vec R$ and are uniformly driven by a linearly polarized laser. Their electronic configuration is composed by three electronic levels of Hydrogen, which consists of the ground state $|1\rangle$ and the two excited states $|2\rangle$ and $|3\rangle$. The transitions $|1\rangle\to|2\rangle$ and $|1\rangle\to|3\rangle$ are parallel optical dipoles with moments $\vec D_{12}^\alpha$ and $\vec D_{13}^\alpha$, respectively, the transition frequencies are denoted by $\omega_{12}$ and $\omega_{13}$ (from now on $\bar{\omega}_{1e}=\omega_{1e}>0$ with $e=2,3$). The reduced level structure allows us to highlight the effects of multilevel interference. Despite the fact we consider the parameters of two transitions of the Hydrogen atoms, however, the choice we perform breaks the rotational symmetry of the atoms. This shall be kept in mind when discussing the single-emitter properties.

The dynamics induced by the laser is described by a Hamiltonian term, which is added to the Hamilton operator of Eq. \eqref{eq:ME}. This procedure corresponds to assuming that the laser field is described by a coherent state and to moving to the reference frame where the quantum state of the laser field is in the vacuum \cite{Giannelli:2019}. We denote by $\omega_L$ the laser frequency, and assume that the laser polarization is linear and that the spatial dependence of the laser field wave vector $\vec k_L$ is orthogonal to the vector $\vec{R}$ joining the two emitters. The laser-atom Hamiltonian has the form
 \begin{equation}
  \hat{H}_L=-\hbar \sum_{\alpha=1,2} \sum_{e=2,3} 
  	g_{1e}^{\alpha} e^{-{\rm i}\omega_Lt}  \hat{\zeta}_{1e}^{\alpha \dagger}+ \text{H.c.}\,,
  \end{equation}
where we have introduced the Rabi frequency $g_{1e}^{\alpha}=-\vec{d}_{1e}^{\alpha}\cdot \vec{E}_L/2\hbar$, which depends on the electric field amplitude $\vec{E}_L$. The master equation takes the form
\begin{equation}
\label{eq:working_form} 
\partial_t\hat \rho=\frac{1}{{\rm i}\hbar}[\hat H_A+\hat H_S,\hat \rho(t)]+\mathcal{L}_{D}\hat \rho(t)+\frac{1}{{\rm i}\hbar}[ \hat{H}_L,\hat \rho]\,,
\end{equation}
where now the sums over the atoms run to $N=2$ and the sums over the internal transitions include just the two transitions with parallel dipolar moments. For simplicity, thus, we can now replace the sum over the transitions $i={i_1,i_2}$ with the sum over the excited state $e=2,3$. Using the simplified level structure we simplify the Hamiltonian term $\hat H_{12}^S$, Eq. \eqref{eq:Lamb+-}, as follows:
\begin{equation} \label{eq:working_form:H:S} 
\hat H_{12}^S=-  \sum_{e,e'=2}^{3}
\mathcal{F}_c(\omega_{1e}+\omega_{1e'})\Xi_{e\,e'}^F(\vec R)\hat{\zeta}_{1e}^{1 \dagger } \hat{\zeta}_{1e'}^{2} +{\rm H.c.}\,,
\end{equation}
where $\mathcal{F}_c(\omega_{1e}+\omega_{1e'})$ is obtained by means of a smoothening of the fast-oscillating function $\Theta_{ij}^{(\Delta t)}$, see Ref. \cite{Buchheit:2016} and Sec. \ref{Sec:CG}, and
\begin{eqnarray}
&&\Xi_{e\,e'}^F(\vec R) =
  \vec{D}_{1e}\cdot\vec{D}_{1e'}\left(\frac{\omega_{e\,e'}}{c}\right)^3\left(  y_0(kR) -  \frac{y_1(kR)}{kR}\right)\,.
\end{eqnarray}
Here we used that the atomic dipole moments are real vectors and introduced the notation $\omega_{e\,e'}=(\omega_{1e}+\omega_{1e'})/2$. Moreover, we have used that the dipole moments are orthogonal to the vector connecting the two atoms. When the interference between different transitions is discarded, then $\mathcal{F}_c(\omega_{1e}+\omega_{1e'})=\delta_{e,e'}$ and  this term takes  the form of the collective Lamb shift of Ref. \cite{James:1993} for the corresponding laser excitation. 

In the dissipator's coefficient we also use the smoothening procedure by replacing $\Theta_{ij}^{(\Delta t)}$ with $\mathcal{F}_c(\omega_{i}+\omega_{j})$. Moreover, we discard the temperature-dependent terms since they give negligibly small contribution at $T=300$ K and optical frequencies. 

\subsection{Photon-count signal}

In order to study the effect of multilevel interference we determine the excitation spectrum $S(\delta_L)$ over the whole solid angle and as a function of the laser detuning $\delta_L=\omega_L-\omega_{12}$. The excitation spectrum (or photon count signal) is defined as:
 \begin{equation} \label{eq:photon_signal}
  	S(\delta_L) =  \sum_{\alpha,\beta}\sum_{e,e'} \Gamma_{\alpha\beta}^{e\,e'\,F}  \text{Tr}[\hat{\zeta}_{1e'}^{\beta}  \hat{\rho}_{\rm st}\hat{\zeta}_{1e}^{\alpha\,\dagger}]\,,
  \end{equation} 
and it is calculated for the steady-state density matrix $\hat{\rho}_{\rm st}$, which is the solution of Eq. \eqref{eq:ME} at eigenvalue zero, $\partial_t\hat\rho_{\rm st}=0$. In our simulations we take the parameters of the transition 2S$\to$4P of Hydrogen. Specifically the ground state is $|1\rangle=|2s_{\frac{1}{2}}, F=0, M_F=0 \rangle$, the excited states are
$|2\rangle= |4p_{\frac{1}{2}}, F=1, M_F=0 \rangle$ and $|3\rangle= |4p_{\frac{3}{2}}, F=1, M_F=0 \rangle$ and are illustrated in Fig. \ref{fig:Level_scheme}. Further details of the parameters are given in Appendix \ref{App:B}. The coefficients are calculated taking a coarse grained time scale $\Delta t=10^{-11}$ sec (see Sec. \ref{Sec:CG} for the analysis of the dependence of the results on the choice of the coarse-graining time scale). For further details we refer the reader to the discussion at the end of this section. We note that, for the level scheme which breaks rotational symmetry, the excitation spectrum of a single emitter exhibits non-vanishing shifts even after integration over the whole solid angle \cite{Buchheit:2016}.

\begin{figure}
  	\includegraphics[width=0.8\linewidth]{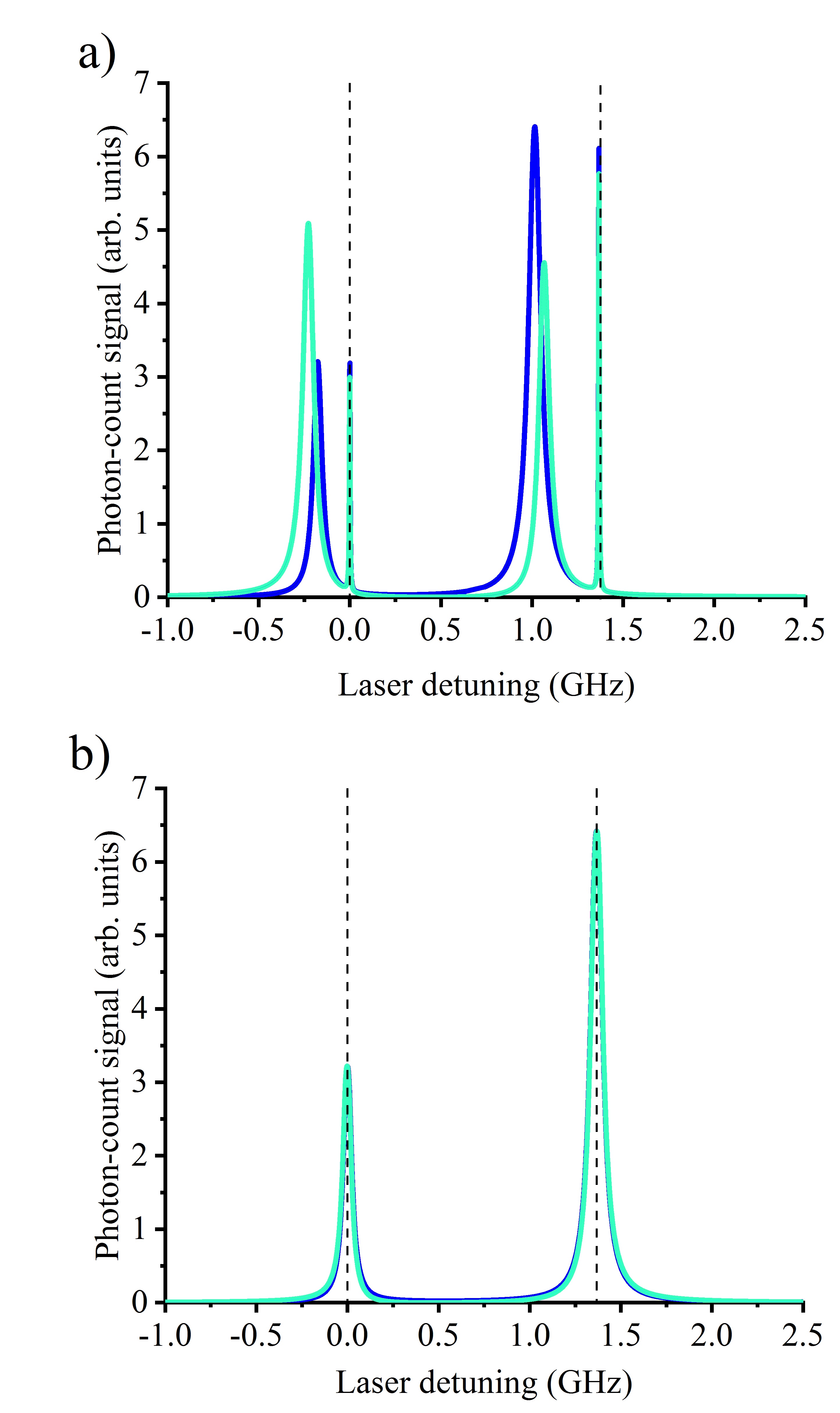}
  	\caption{\label{fig:curve}
 Photon count signal, Eq. \eqref{eq:photon_signal}, for two emitters as a function of the laser detuning $\delta_L$ and at interatomic distance (a) $R=0.01 \, \mu$m and (b) $R=0.1 \, \mu$m. The cyan (light gray) curve is calculated with the full master equation Eq. \eqref{eq:ME}; The blue (dark gray) curve is calculated by setting all cross-interference terms to zero in Eq. \eqref{eq:ME}.  The Rabi frequency for the $|1\rangle\to|3\rangle$ transition is $g_{13}=20 \, \gamma_{3}$, where $\gamma_3$  is the decay rate from the state $4P_{3/2}^{F=1}$ to the state $2S_{1/2}^{F=0}$. 
The coarse-graining time is taken to be $\Delta t=10^{-11}$ sec. The vertical dashed lines indicate the frequency $\omega_{12}$ and $\omega_{13}$ of the individual atomic resonances. The parameters of the atomic transitions are reported in the text and in Appendix \ref{App:B}.}
 \end{figure} 	

Figure \ref{fig:curve} displays the photon count signal (cyan line) for a given value of the laser intensity and as a function of the laser detuning $\delta_L$ for two interatomic distances: (a) $R=0.01 \, \mu {\rm m}$  and (b) $R=0.1 \, \mu {\rm m}$. These shall be compared with the wavelength $\lambda_{12}=2\pi c/\omega_{12}=0.468\, \mu m$ such that (a) corresponds to $k R\simeq 0.13$ and (b) to $kR\simeq 1.3$. The orange line gives the signal obtained when one artificially sets the multilevel interference effects to zero (corresponding to setting $\Theta_{ij}^{(\Delta t)}\to \delta(\bar\omega_i-\bar\omega_j)$, namely $\Delta t\to 0$). The mismatch between the cyan and the orange superradiant peaks is caused by the cross-interference terms.

We start with discussing the case $R=0.1 \, \mu {\rm m}$, when the interatomic distance is of the order of the wavelength. In this case the photon count signal is dominated by the photon count signal of the individual atoms, the peak maxima are at the frequency of the atomic levels, there are no evident features which could be attributed to superradiance and/or subradiance. Here, the inclusion of cross-interference terms gives rise to a slightly visible discrepancy between the two curves in the frequency interval between the two peaks.  When decreasing the interatomic distance to $R=0.01 \, \mu {\rm m}$ the spectroscopic lines are splitted into the sub- and superradiant components. The frequency gap between the peaks of the sub- and superradiant components is given by the corresponding diagonal frequency shifts of Eq.  \eqref{H:Lamb}. In the next section we determine the line shifts one extracts by analysing these spectra.

\subsection{Line shifts due to cross interference}

In order to quantify the effect of the cross-interference terms, we determine the line shifts $\delta\omega_{j}$ due to the multilevel interference. We focus on the lines of the superradiant states and extract the shift 
	\begin{equation}
    	\delta\omega_{j}= \frac{1}{2\pi}\left(x_j^\prime-x_j\right)\,,
    	\end{equation}
where the quantity $x_j^\prime$  (with $j=2,3$ for $|1\rangle \to |j\rangle $) is extracted from the photon-count signal calculated using the master equation \eqref{eq:ME}. The frequency $x_j$, instead, is obtained by artificially setting all multilevel interference terms to zero, namely, by setting $\Theta_{ij}^{(\Delta t)}\to\delta(\bar{\omega}_i-\bar\omega_j)$ in the coefficients of Eq. \eqref{eq:ME}. Thus, the frequency $x_j$ includes also the collective Lamb shift.  The line shifts we report are determined  from the photon count signal as a function of the interatomic distance by taking the limit of vanishing Rabi frequencies, and are extracted by fitting the photon count signal using the following function, which is the sum of two Lorentzian curves:
    \begin{multline} \label{eq:2lorentz}
    S^{LL}(x) = \frac{a_2}{\pi} \frac{b_2/2}{(x-x_2)^2+(b_2/2)^2} +
    \\
    + \frac{a_3}{\pi} \frac{b_3/2}{(x-\omega_0-x_3)^2+(b_3/2)^2}\,,
    \end{multline} 
and $\omega_0=2\pi\nu_0$ is the frequency gap between state $|2\rangle$ and $|3\rangle$ and is given in Appendix \ref{App:B}. It discards the presence of the subradiant peaks, whose magnitude becomes very small at low Rabi frequencies (for instance, for Rabi frequencies that are 1\% the natural linewidth the magnitude is approximately $10^{-3},10^{-4}$ smaller than the superradiant ones). Nevertheless, these signals are generally different from zero and give rise to a systematic error in determining the line shift of the superradiant resonance. We remark that the choice of the fitting function is not optimal: {In fact, Eq. \eqref{eq:2lorentz} corresponds to the spectroscopic signal due to the sum of two independent decay processes, and does not properly catch the features due to interference.}
Indeed, the data in Fig. \ref{fig:curve} shows that the curves are more similar to Fano-like profiles. Previous studies showed that the excitation spectra of optically dense (homogeneously-broadened) media differ from Lorentz resonances \cite{James:1993,Putnam:2016,Zhu:2016}. Our choice is thus not going to be a reliable estimate of the shifts induced by multilevel interference. We expect, nevertheless, that it allows us to gain insight into their order of magnitude. 

\begin{figure}
   	\includegraphics[width=1\linewidth]{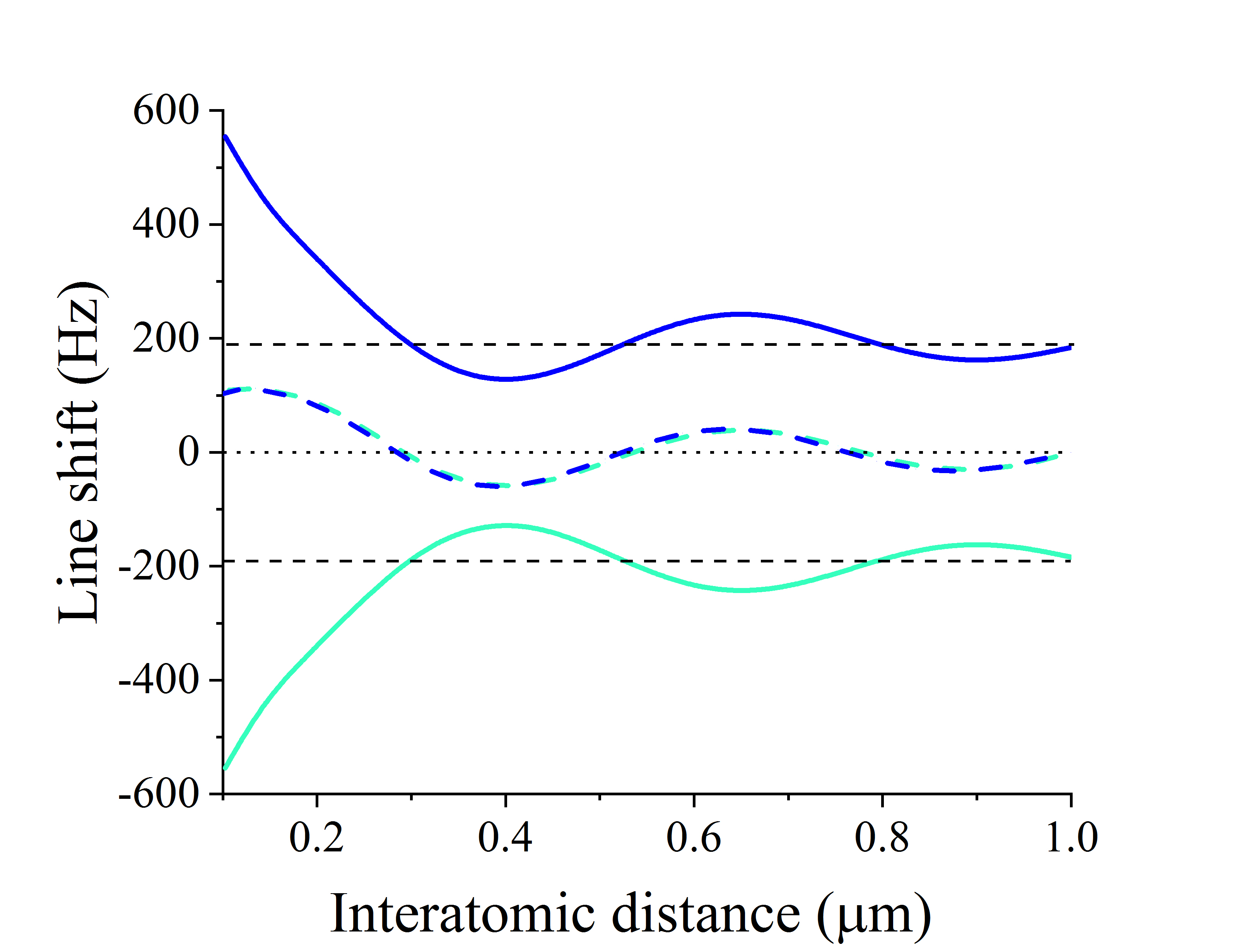}
   	\caption{\label{fig:all_included}  Line shifts versus the interatomic distance for two emitters transversally driven by linearly-polarized light. The cyan (light grey) curve corresponds to the line of transition $|1 \rangle \rightarrow |2 \rangle $ and the blue (dark grey) ones to the transition ($|1 \rangle \rightarrow |3 \rangle $). The solid lines are extracted from the photon count signal using the fit of Eq. \eqref{eq:2lorentz}, the dashed lines are the curves in the absence of multilevel interference, where ideally $\delta\omega_j=0$. The deviation from zero is here due to the fact that we have discarded the presence of the subradiant peaks in applying the fitting function \eqref{eq:2lorentz}. The horizontal dotted lines indicate the shift due to multilevel interference in a single three-level emitter ("atom"). }
   \end{figure}

Figure \ref{fig:all_included} shows the line shifts as a function of the interatomic distance: at sufficiently short distances the shifts are significantly larger than the ones predicted for a single emitter and above the systematic error, due to discarding the subradiant peaks and illustrated by the dashed lines. The line shifts tend to increase the frequency gap between the two excited states as $R\to 0$, while for $R\to\infty$ they converge to the values indicated by the dashed lines, that are the shifts we calculate for the case of a single artificial emitter composed by three levels. 

We now argue that the observed shifts are due to quantum interference between the processes illustrated in Fig. \ref{Fig:1}. For this purpose we analyse the shifts by considering two artificial cases:  (i) The {\it single-atom cross interference}, in which  we only consider the scattering processes displayed in Figures \ref{Fig:1}a) and \ref{Fig:1}b). This corresponds to set $\Delta_{ij}^{12}=0$ in  \eqref{eq:Lamb+-} and $\Gamma^{ij}_{12}=0$ in  \eqref{eq:dissipator} for $i\neq j$. (ii) The {\it interatomic cross interference}, in which we discard  scattering processes displayed in Figure \ref{Fig:1}a) and we keep the others. In this case we set $\Delta_{ij}^{\alpha\alpha}=0$ and $\Gamma^{ij}_{\alpha\alpha}=0$ for $i\neq j$.  We further separately analyse the effect of the cross-damping terms (namely, the terms of the master equations where multilevel interference appears in the dissipator) and of the cross-shift terms (where multilevel interference appears in the Hamiltonian \eqref{eq:Lamb+-}).

\begin{figure}
   	\includegraphics[width=0.8\linewidth]{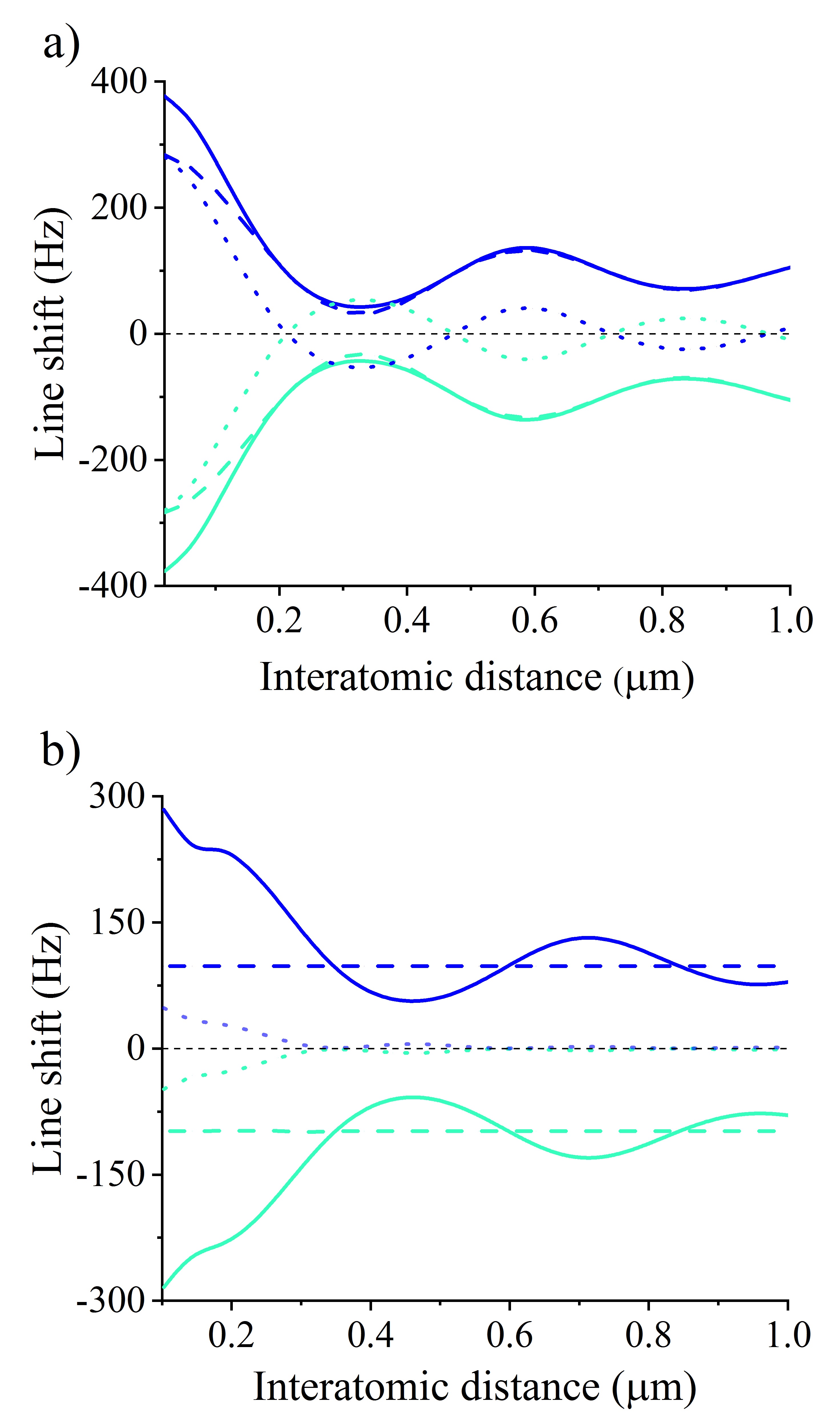}
   	\caption{\label{fig:individual_interference_contribution}
 Line shifts versus the interatomic distance for two atoms transversally driven by a linearly polarized laser due to (a) the cross-damping terms (after setting all cross shift terms $\Delta_{ij}^{\alpha\beta}=0$ for all $i\ne j$ in Hamiltonian \eqref{H:Lamb}) and (b) the cross shift terms (after setting all cross damping terms $\Gamma_{ij}^{\alpha\beta}=0$ for all $i\ne j$ in the dissipator \eqref{eq:LD}).   The cyan (light grey) curves correspond to the line of transition $|1 \rangle \rightarrow |2 \rangle $ and the blue (dark grey) ones to the transition $|1 \rangle \rightarrow |3 \rangle $). The dashed curves correspond to case (i),  the dotted lines correspond to the case (ii), the solid lines include both intratomic and interatomic cross damping (a) and cross-shift (b) terms. }
   \end{figure}

We first study the impact of the cross-damping terms versus $R$ and artificially set all terms $\Delta_{ij}^{\alpha\beta}=0$ with $i\ne j$ in Hamiltonian \eqref{H:Lamb}.
Figure \ref{fig:individual_interference_contribution}a) represents the results when we include  the cross-damping terms (i) only in the single-atom dissipator (intratomic, $\alpha=\beta$), (ii) only in the interatomic dissipator (interatomic, $\alpha\neq\beta$) and (iii) when we consider both intratomic and interatomic cross-damping terms. In the case (i) the shifts due to the {\it single-atom} cross-damping terms at large distance oscillate around a magnitude of $\sim 100$ Hz. In the case (ii) the line shifts vanish for $R\to\infty$. For vanishing distances the line shifts (i) and (ii) converge to a similar value. The total contribution of the intra- and interatomic cross-damping terms is not additive, as visible when comparing these curves with the ones obtained including both kinds of cross-damping terms.  
Figure \ref{fig:individual_interference_contribution}b) displays the impact of the cross-shift terms on the line shifts after artificially setting all terms $\Gamma_{ij}^{12}=0$ in the dissipator \eqref{eq:LD}. Over  the interval of distances $R=[0.1,1] \mu {\rm m}$.  The total line shift has some oscillatory behaviour which tends to the single-atom result as $R$ increases. At small $R$ the cross-shift terms become dominant and tend to increase the frequency gap between the spectroscopic lines. 

\begin{figure}
            	\includegraphics[width=0.8\linewidth]{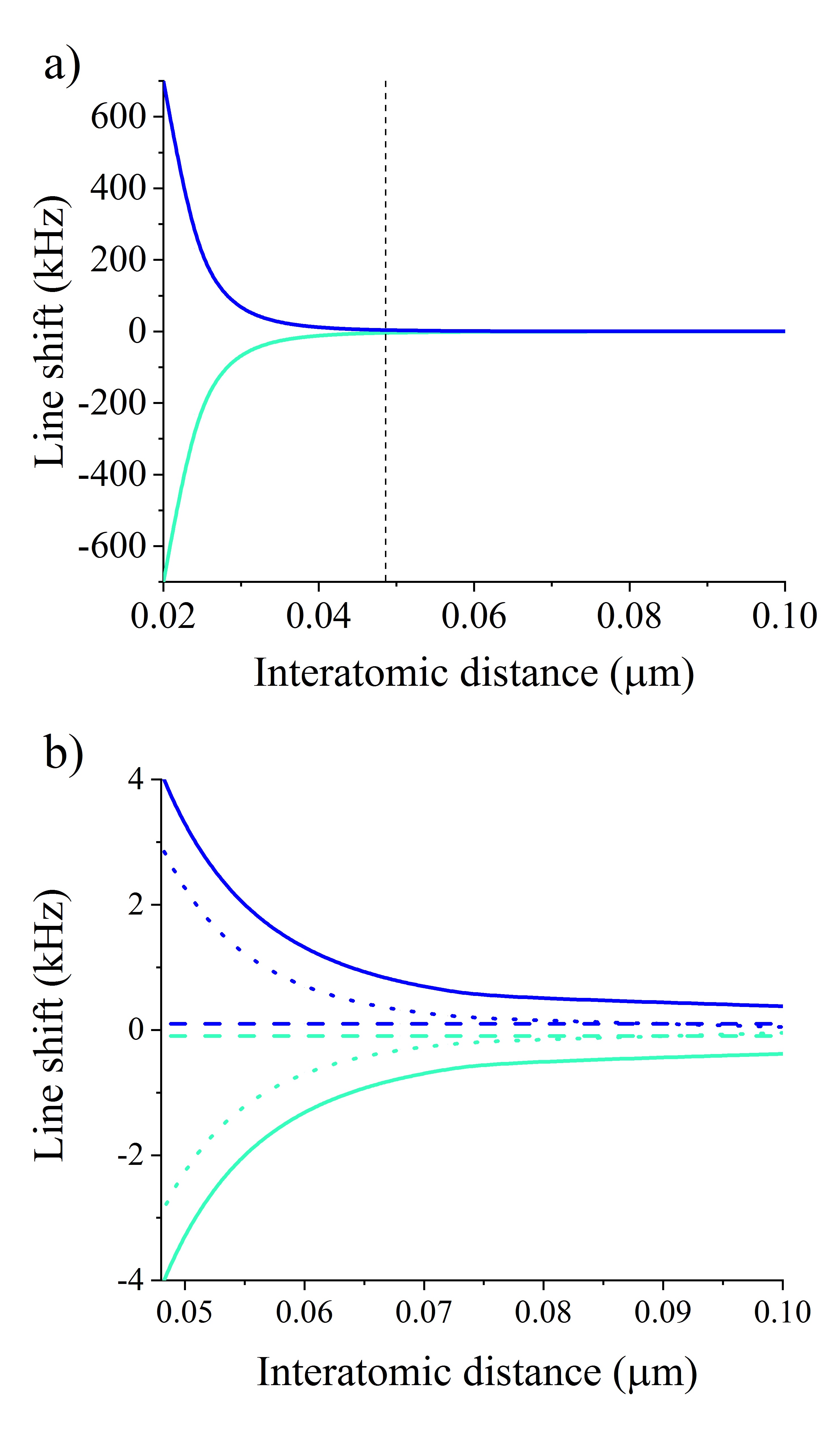}
            	\caption{\label{fig:Short_distance_CST}
Same as Fig. \ref{fig:individual_interference_contribution}b) but for interatomic distances below $\lambda/5$, the vertical dotted line indicates the value $\lambda/10$. Subplot (b) zooms on the behaviour in the interval $[\lambda/10,\lambda/5]$. }
\end{figure}

The  behaviour at short distances is diplayed in Fig. \ref{fig:Short_distance_CST}. Here it is evident that the cross-shift terms are responsible for large shifts of the lines. Below $R=48$ nm (which corresponds to $R\sim\lambda/10$) the shift of the line $|1\rangle\to|3\rangle$ increase rapidly to the magnitude of 0.6 MHz, which starts to be comparable with the natural linewidth for optical transitions. 
 
\subsection{About the coarse graining time scale}
\label{Sec:CG}

The use of the coarse graining master equation allows one to derive {\it ab initio} a master equation fulfilling the Lindlad form and yet systematically including the cross-interference terms. The drawback is the explicit dependence on the coarse graning time, which becomes visible in the  functional form of $\Theta_{ij}^{(\Delta t)}$, Eq. \eqref{Eq:Theta:ij}, and which multiplies all coefficients for $i\neq j$. 
We note that this function determines the frequency window, for which the interference of two parallel dipolar transitions give rise to relevant contributions to the dynamics. 

One striking property is that $\Theta_{ij}^{(\Delta t)}$ gives rise to strong oscillations of the coefficients with $\Delta t$. The oscillations are majorly due to the sharp time intervals over which the dynamics has been divided and could be eliminated by introducing a smoothening, for instance by taking a Gaussian function of width $\Delta t$ and calculating the convolution \cite{Buchheit:2016}
\begin{equation} \label{eq:smoothing}
\Theta_{ij}^{(\Delta t)} \rightarrow \mathcal{F}_c  (\omega_i+\omega_j)= \int_{0}^{\infty} dx \Theta_{ij}(x)\frac{{\rm e}^{-x^2/\Delta t^2}}{\sqrt{\pi\Delta t/2}}\,.
\end{equation} 
This smoothening procedure delivers the new damping coefficients
\begin{eqnarray} \label{eq:Gamma_F}
	\Gamma_{\alpha\beta}^{ij\,\,(F)}=\mathcal{F}_c   (\omega_i+\omega_j)\frac{\vec{D}_{i}^{\alpha*}\cdot\vec{D}_{j}^{\,\beta} }{{2\pi}\hbar c^{3}} \omega_{ij}^{3}  F_{\alpha\beta}^{ij}(k_{ij})\,, 
\end{eqnarray}
which preserves the Lindblad form of the density matrix. Similarly we obtain the cross-coupling Hermitian terms after the smoothening.

\begin{figure}
  	\includegraphics[width=1.0\linewidth]{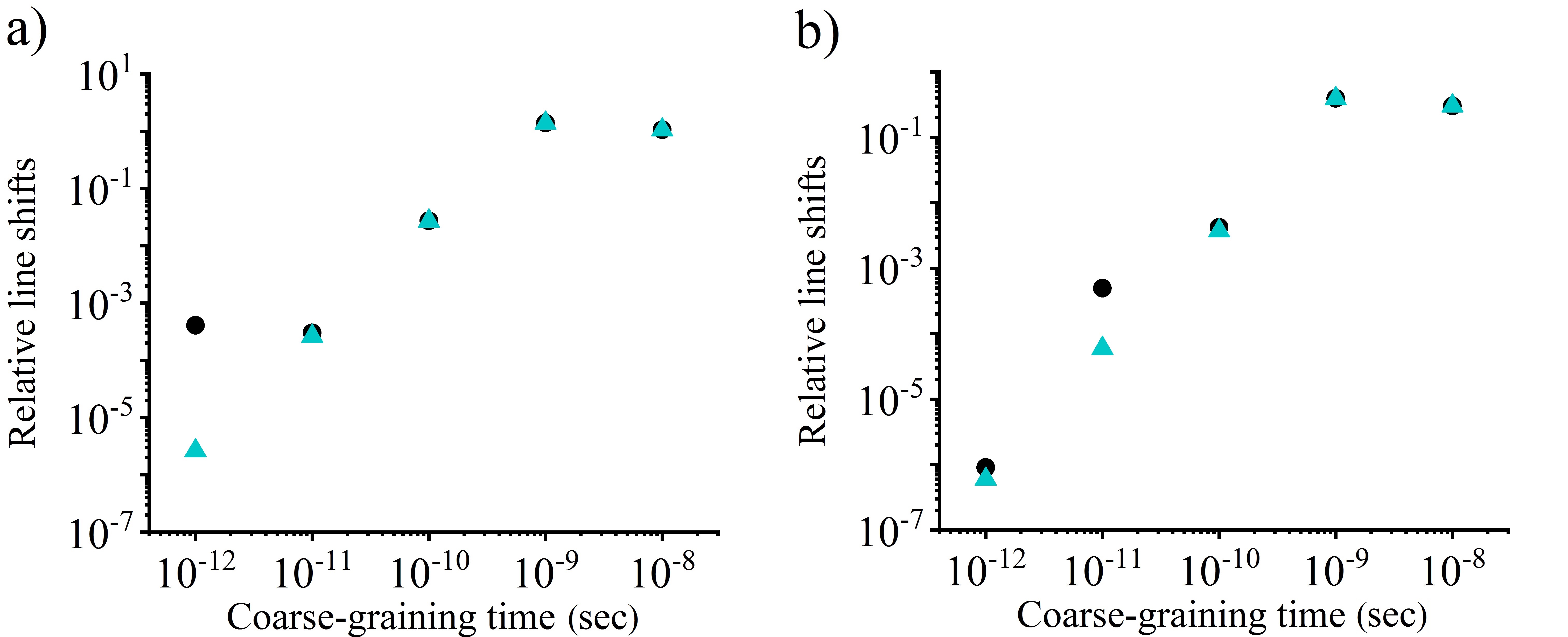}
  	\caption{\label{fig:CG_analysis}
Dependence of the relative line shifts on the coarse-graining parameters. The relative line shifts are defined as $|(x(\Delta t_i)-x(\Delta t_{i+1}))/x(\Delta t_i)|$. Here $\Delta t_i = 10^{-i}$ sec is the coarse-graining time and the index $i$ takes integer values from 8 to 12. Cyan (black) triangles correspond to relative shift of the first  (second) line. (a) corresponds to the interatomic distance $R=0.1 \, \mu$m, (b) to $R=1 \, \mu $m.}
  \end{figure}

Even after this smoothening, the coefficients of the master equation still depend on the choice of $\Delta t$. For the master equation to be valid, their value shall be independent on the specific choice of $\Delta t$ over an interval of value. A rigorous lower bound for $\Delta t$ can be found by imposing the positivity of the Lindblad equation, as discussed in Ref. \cite{Farina:2019}. An heuristic approach is based on identifying the coarse-grained time for which the scattering properties are stable over several orders of magnitude, such that $\tau_R\ll\Delta t$ and $\Delta t$ is smaller than the smallest rate of the system dynamics. Figure \ref{fig:CG_analysis} shows the line shifts for different values of the coarse-graining time. 
The results do not vary over the interval of values of $\Delta t$, over which we expect that the time-scale separation ansatz holds. They start to appreciably vary for $\Delta t > 10^{-10}$ sec, and thus when $\Delta t$ becomes comparable with the natural lifetime of the excited states, which is here of the order of $10^{-8}$ sec.


\section{Conclusions} \label{chapter:conclusions}

In this work we have presented the systematic derivation of a master equation for an optically dense medium, which is composed by multilevel emitters. The master equation fulfils the Lindblad theorem \cite{Breuer} and includes the effect of interference between transitions which have parallel dipoles. This interference is induced by vacuum effects and gives rise to additional terms in the dissipator and Hamiltonian which can mutually interfere and whose strengths depend on the mean interparticle distance. 

We have provided a numerical example where we have applied our master equation to two identical emitters each consisting of  two parallel dipoles with a common ground state. We have shown that, even if the dipoles are not resonant, vacuum induced interference gives rise to measurable effects in the excitation spectrum. We have verified that the magnitude of the shifts depend on the ratio between the frequency gap between the interfering dipoles and their average linewidth and increases as this ratio decreases \cite{Ficek}, they become more evident when the interparticle distance decreases and emerge from the interplay of the interference between parallel dipoles of a single emitter and of the two emitters. Moreover, for realistic configurations the photodetection signal depends on the angle of emission and can be larger for certain directions \cite{Buchheit:2016}. 

Future work shall focus on alkali or alkali-earth metal atoms, consider the full sublevel structure and analyse the spectrum at different detection angles. A more accurate choice of the fitting functions shall provide a better estimate of the line shift due to multi-level interference \cite{Jentschura:1997,Putnam:2016}. This model, moreover, can be extended to Rydberg transitions \cite{Lahaye:2017}, where the multilevel interference is expected to be more prominent \cite{Cardimona:1983}, and to molecules \cite{Zelevinsky:2015}. 

The master equation here derived  {can be extended and applied to studying propagation of quantum light in superradiant media and confined geometries \cite{Zhou:2017,Chen:2018,Zhou:2020}. By means of the input-output formalism \cite{Collett:1985,Carmichael} one can extract from our model the coherence properties of the scattered light and analyse the effect of vacuum induced interference on field- and intensity-intensity correlation functions. } Future studies will analyse its prediction  on light transport in a disordered medium \cite{Javanainen:2014,Zhu:2016, Cottier:2018} and in an ordered array of emitters \cite{Longo:2015,Facchinetti:2016} for level configurations where vacuum-induced interference is expected to be relevant.

\acknowledgements The authos are grateful to Guido Pupillo, Johannes Schachenmeyer, and to the ITN Network ColOpt members for scientific discussions. We thank Andreas Buchheit for help in the first stages of this project and Anette Messinger for discussions and for careful reading of this manuscript. Funding by the EU ITN Network ColOpt and by the German Research Foundation (DFG, Priority Programme No. 1929, GiRyd) is gratefully acknowledged.

\begin{appendix}
\section{Derivation of the Born-Markov master equation in the coarse graining formalism}
\label{App:MEq}

The second integrand on the right-hand side of Eq. \eqref{eq:decomposition} is here reported after applying the Born approximation:
\begin{eqnarray}
\Lambda_2^{\alpha,\beta }(T,\tau)\tilde{\rho}(\tau_-)&=&\sum_{i,j}\bar C_{ij}^{\alpha\beta}(\tau)\left(\left[{\tilde{\sigma}}_{j}^{\beta}(\tau_-) \tilde{\rho}(\tau_-), {\tilde{\sigma}}_{i}^{\alpha}(\tau_+)\right]\right.
 \nonumber\\
&&\left.+\left[{\tilde{\sigma}}_{i}^{\alpha}(\tau_+),\tilde{\rho}(\tau_-){\tilde{\sigma}}_{j}^{\beta}(\tau_-)\right]\right) +\text{H.c.} \,,
 \label{eq:map:0}
 \end{eqnarray}
where $\tau_\pm=T\pm\tau$. Subscript $i$ labels a pair of level coupled by a non-vanishing dipole moment: $i\equiv i_1,i_2$ with dipole moment $\vec{d}_{i}^{\alpha}=\, _\alpha\langle i_1|\vec{d}|i_2\rangle_\alpha$. The function $\bar C_{ij}^{\alpha\beta}(\tau)$ specifically reads
\begin{eqnarray}
\bar C_{ij}^{\alpha\beta}(\tau)&=&\sum_{\lambda}\left(g_i^\lambda \bar g_j^\lambda (n(\omega_\lambda,T)+1){\rm e}^{-{\rm i}\omega_\lambda \tau}{\rm e}^{{\rm i}{\vec k_\lambda}\cdot(\vec R_{\alpha}-\vec R_{\beta})}\right.\nonumber\\
& &\left.+\bar g_i^\lambda  g_j^\lambda n(\omega_\lambda,T){\rm e}^{{\rm i}\omega_\lambda \tau}{\rm e}^{-{\rm i}{\vec k_\lambda}\cdot(\vec R_{\alpha}-\vec R_{\beta})}\right)\,,
\end{eqnarray}
where $n(\omega,T)=1/[\exp(\hbar\omega/k_BT)-1]$ is the mean photon number at frequency $\omega$ and temperature $T$ and the sum over the modes is bounded by the cutoff frequency $\omega_{cut}$. In the continuum limit it is given by the expression
\begin{eqnarray}
\label{eq:correlation_function}
\bar C_{ij}^{\alpha\beta}(\tau)&\to&\int\limits_0^{\omega_{cut}} \frac{d \omega}{(2\pi)^2\hbar c^3}\, \omega^{3} \left([1+ n(\omega,T) ]e^{-i\omega\tau} \right. \nonumber\\
 & & \left.+ n(\omega,T) e^{i\omega\tau}\right)  \bar F^{ij}(k,\vec{R}_{\alpha\beta})\,.
\end{eqnarray}

Assuming the Born-Markov approximation, we can write $\tilde{\rho}(T-\tau)\approx \tilde{\rho}(T)$ in Eq. \eqref{eq:map:0} \cite{Breuer,Lidar:2001}. We also note that, consistently with the Markov approximation, $\tilde{\rho}(T)$ is essentially constant over the interval of integration $[t,t+\Delta t]$ of the variable $T$. We then set $\tilde{\rho}(T)= \tilde{\rho}(\bar t)$ with $\bar t=t+\Delta t/2$. Using that $\tilde\sigma_j^\beta(t_1)={\rm e}^{{\rm i}\omega_j(t_1-\bar t)}\tilde\sigma_j^\beta(\bar t)$, we first rewrite Eq. \eqref{eq:map:0} as
\begin{eqnarray}
\Lambda_2^{\alpha,\beta }(T,\tau)\tilde{\rho}(\tau_-)&\approx&\sum_{i,j}\mathcal C_{ij}^{\alpha\beta}(T,\tau)
\left(\left[{\tilde{\sigma}}_{j}^{\beta}(\bar t) \tilde{\rho}(\bar t), {\tilde{\sigma}}_{i}^{\alpha}(\bar t)\right]\right.
 \nonumber\\
&&\left.+\left[{\tilde{\sigma}}_{i}^{\alpha}(\bar t),\tilde{\rho}(\bar t){\tilde{\sigma}}_{j}^{\beta}(\bar t)\right]\right) +\text{H.c.} \,,\end{eqnarray}
where 
\begin{equation}
\mathcal C_{ij}^{\alpha\beta}(T,\tau)=\bar C_{ij}^{\alpha\beta}(\tau){\rm e}^{{\rm i}(\omega_i+\omega_j)T}{\rm e}^{{\rm i}(\omega_i-\omega_j)\tau}.
\end{equation}
Using now that $\tilde{\sigma}^\alpha_j(t)=\exp({\rm i}\omega_jt)\sigma_j^\alpha$ in Eq. \eqref{eq:map:0}, the time integrals take the form:
\begin{eqnarray}
&&\frac{1}{2\Delta t} \int_{-\Delta t}^{\Delta t} dT{\rm e}^{\pm i(\omega_i-\omega_j) T/2} \int_{-\Delta t}^{\Delta t}  d\tau \, \theta(\tau)C_{\alpha\beta}(\tau) {\rm e}^{\pm i(\omega_i+\omega_j) \tau/2}\nonumber\\
&&=\Theta_{ij}^{(\Delta t)}\int_{-\Delta t}^{\Delta t}  d\tau \, \theta(\tau)C_{\alpha\beta}(\tau) {\rm e}^{\pm i(\omega_i-\omega_j) \tau/2}\,,
\end{eqnarray}
where 
\begin{equation}
\Theta_{ij}^{(\Delta t)}=\frac{\sin((\omega_i+\omega_j)\Delta t/2)}{(\omega_i+\omega_j)\Delta t/2}\,.
\end{equation}
When the transition are in the optical range, this function selects secular terms. For this reason, in the following we restrict the sum to all pairs such that $\omega_i>0$. 
The second integral is evaluated after approximating the extrema of integration by $[-\Delta t,\Delta t]\to[-\infty,\infty]$, which is consistent with the assumption that $C(\tau)$ decays to zero over time scales much shorter than $\Delta t$.

\section{Parameters of the simulation}
\label{App:B}

The magnitude of the fine structure splitting for 4p state is taken to be $\nu_0\approx$1.367 GHz \cite{Kolachevsky,Udem:2019} and includes also the Hyperfine structure splitting and QED corrections. We neglect thermal effects: we set $n(\omega_{1e})=0$, which is a good approximation at room temperature $T=300\,K$. Moreover, we take the following values for the radiative shifts: $\Delta_{22}^S=-2\pi\times 1401.52$ kHz for the state $4p_{\frac{1}{2}}$ and $\Delta_{33}^S=2\pi \times 1767.30$ kHz for the state $4p_{\frac{3}{2}}$ \cite{Jentschura:1997}. We then construct the atomic cross-shift term between the excited states using relation \eqref{eq:CST}: $\Delta_{23}^S$= $\Delta_{32}^S=2\pi \times 366.2$ kHz using the relation between dipole moments of the corresponding transitions: $d_{12}=(1/3)d_R$, $d_{13}=-(\sqrt{2}/3)d_R$, where $d_R$ is the radial integral $d_R=\langle 2s|r|4p \rangle = 1.28\,[a.u.]$. The values for the natural line width are  $\gamma_2=\Gamma_{22} = 2\pi\times 511$ kHz for the state $4p_{\frac{1}{2}}$ and $\gamma_3\equiv\Gamma_{33} =2\pi\times 1022$ kHz for the state $4p_{\frac{3}{2}}$. All the cross-interference terms both for the dissipator and the Lamb shift were computed with the coarse-graining time $\Delta t = 10^{-11}$ sec. The computational checks showed that the solutions of the master equation for the chosen system remains stable in this coarse-graining time region, see the Figure \ref{fig:CG_analysis} in Sec. \ref{Sec:CG}.    

\section{Determination of the line shifts}
\label{App:C}

{\it Line shifts due to cross-interference for a single emitter}
In this appendix we illustrate the procedure we apply in order to determine the line shifts due to the multilevel quantum interference terms. We provide the example of a single emitter, and we refer to it using the wording "single atom". However, due to the special structure we assume the emitter is not rotationally invariant, which changes the spectroscopic properties and gives rise to a global line shift when integrating the photon count signal over the whole solid angle. 

The line shifts for a single atom are defined as  
 \begin{equation} \label{single-atom_line_shift_nointerference}
 	\Delta_j(g) = \frac{1}{2\pi}\left(x_j^{g} - x_j^{\text{eigen}}\right),
 \end{equation}    
where $x_j^{\text{eigen}}$ is the eigenfrequency of the $j^{th}$ transition  ($j=1$ for $|1\rangle \to |2\rangle $ and $j=2$ for $|1\rangle \to |3\rangle $) which we extract from the master equation when we set all multilevel interference terms to zero, and $x_j^{g}$ is the line position obtained by fitting the photon count signal using the fitting function  \eqref{eq:2lorentz}. The line shifts $\Delta_j(g)$ depends on the laser intensity and thus on the Rabi frequency $g$.  The line shift we identify corresponds to the limit:
 \begin{equation} \label{single-atom_line_shift_nointerference:limit}
 	\Delta_j =\lim_{g\to 0}\Delta_j(g).
 \end{equation}    
 \begin{figure}
  	\includegraphics[width=1.0\linewidth]{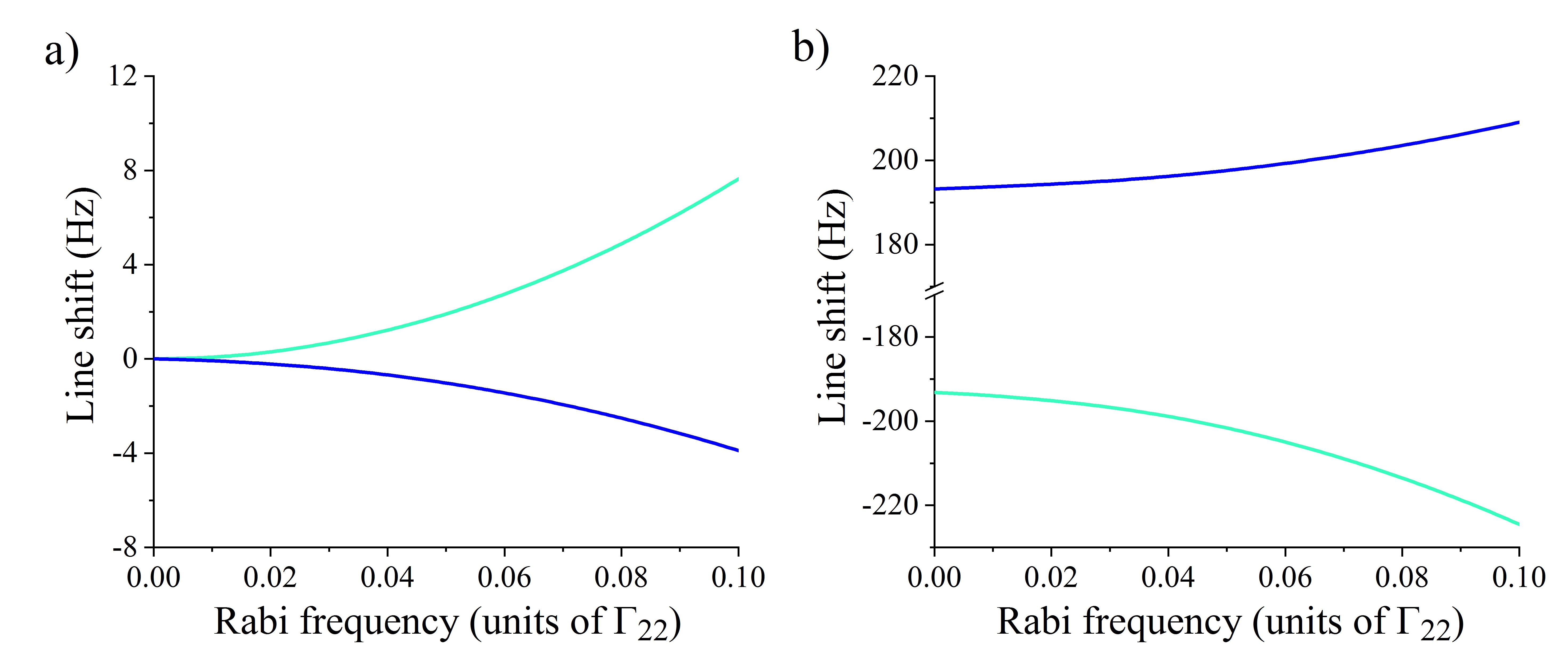}
  	\caption{\label{single_atom} 
 Line shift $\Delta_j(g)$ of a single emitter as a function of the Rabi frequency $g_{13}$. The cyan (light grey) curves correspond to the line of transition $|1 \rangle \rightarrow |2 \rangle $ and the blue (dark grey) ones to the transition $|1 \rangle \rightarrow |3 \rangle $). The line shift is extracted from the photon count signal by using the fitting functions of Eq. \eqref{eq:2lorentz}.  Subplot (a) displays the line shifts without any cross-interference terms. In subplot (b) the line shifts are obtained for the full master equation \eqref{eq:ME} for a single emitter. The levels are illustrated in Fig. \ref{fig:Level_scheme}, the parameters are detailed in the text.}
  \end{figure} 
 The limit is extracted from our numerical analysis: We evaluate it for decreasing values of $g$. We report the behaviour in Fig. \ref{single_atom} a) when we set to zero the multilevel interference terms and b) for the full master equation.  The presence of the interference terms shifts both peaks to the magnitudes $\pm$ 195 Hz for vanishing laser intensity.

\end{appendix}

\end{document}